\documentclass[10pt,conference]{IEEEtran}
\IEEEoverridecommandlockouts

\usepackage[most]{tcolorbox}
\usepackage{listings}
\usepackage{float}
\usepackage{booktabs} 
\usepackage[flushleft]{threeparttable}
\usepackage{xcolor}
\usepackage[utf8]{inputenc}
\usepackage{ifthen}
\usepackage{multirow}
\usepackage{caption}
\usepackage{hyperref}
\hypersetup{colorlinks=true}
\usepackage{url,moreverb,xspace}
\usepackage{enumitem}
\usepackage{array,graphicx}
\usepackage{balance}
\usepackage{pifont}
\usepackage{etoolbox,siunitx}
\usepackage{nicefrac}
\usepackage{mathtools}
\usepackage{amsmath,amssymb}
\usepackage{mathrsfs}
\usepackage{makecell}
\usepackage{soul}
\usepackage[T1]{fontenc}

\usepackage[labelsep=quad,indention=10pt]{subfig}
\usepackage{xcolor,pifont}
\newcommand*\colourcheck[1]{%
  \expandafter\newcommand\csname #1check\endcsname{\textcolor{#1}{\ding{52}}}%
}
\newcommand*\colourcross[1]{%
  \expandafter\newcommand\csname #1cross\endcsname{\textcolor{#1}{\ding{56}}}%
}

\usepackage[lined,boxruled,norelsize,linesnumbered,noend]{algorithm2e}
\def\HiLi{\leavevmode\rlap{\hbox to \hsize{\color{gray!35}\leaders\hrule height .8\baselineskip depth .5ex\hfill}}}

\newcommand{\crawljax}{Crawljax\xspace} %
\newcommand{\tool}{\textsc{WebEmbed}\xspace} %

\newcommand{\changed}[1]{\textcolor{black}{#1}}
\newcommand*\rot{\rotatebox{90}}

\newboolean{showcomments}
\setboolean{showcomments}{true}
\ifthenelse{\boolean{showcomments}}
{\newcommand{\nb}[2] {
  \fcolorbox{black}{gray!20}{\bfseries\sffamily\scriptsize#1:}
  {\sf\small$\blacktriangleright$\textit{#2}$\blacktriangleleft$}
}
}
{\newcommand{\nb}[2]{}
}

\newcommand{\head}[1]{\noindent\textbf{#1.}}

\definecolor{lightgray}{gray}{0.85}
\definecolor{bananayellow}{rgb}{1.0, 0.88, 0.21}

\begin{document}

\pagestyle{plain}

\title{Neural Embeddings for Web Testing}

\author{
\IEEEauthorblockN{Kasun Kanaththage\textsuperscript{1}, 
Luigi Libero Lucio Starace\textsuperscript{2}, 
Matteo Biagiola\textsuperscript{3,4}, 
Paolo Tonella\textsuperscript{4}, 
Andrea Stocco\textsuperscript{1}}
\IEEEauthorblockA{\textsuperscript{1}\textit{Technical University of Munich}, Munich, Germany\\
\{kasun.kanaththage, andrea.stocco\}@tum.de}
\IEEEauthorblockA{\textsuperscript{2}\textit{Università degli Studi di Napoli Federico II}, Napoli, Italy\\
luigiliberolucio.starace@unina.it}
\IEEEauthorblockA{\textsuperscript{3}\textit{University of St.~Gallen}, St.~Gallen, Switzerland\\
matteo.biagiola@usi.ch}
\IEEEauthorblockA{\textsuperscript{4}\textit{Università della Svizzera italiana}, Lugano, Switzerland\\
\{matteo.biagiola, paolo.tonella\}@usi.ch}
}

\IEEEtitleabstractindextext{%
\begin{abstract}
Web test automation techniques often rely on crawlers to infer models of web applications for automated test generation. However, current crawlers rely on state equivalence algorithms that struggle to distinguish near-duplicate pages, often leading to redundant test cases and incomplete coverage of application functionality. 
In this paper, we present a model-based test generation approach that employs transformer-based Siamese neural networks (SNNs) to infer web application models more accurately. By learning similarity-based representations, SNNs capture structural and textual relationships among web pages, improving near-duplicate detection during crawling and enhancing the quality of inferred models, and thus, the effectiveness of generated test suites.
Our evaluation across nine web apps shows that SNNs outperform state-of-the-art techniques in near-duplicate detection, resulting in superior web app models with an average $F_1$ score improvement of 56\%. These enhanced models enable the generation of more effective test suites that achieve higher code coverage, with improvements ranging from 6\% to 21\% and averaging at 12\%.
\end{abstract}

}

\maketitle

\IEEEdisplaynontitleabstractindextext

%
\IEEEpeerreviewmaketitle

\section{Introduction}\label{sec:introduction}

Test automation enables end-to-end (E2E) functional testing of web applications by simulating user interactions through automated scripts~\cite{ricca2019neonate,2016-leotta-Advances,DBLP:journals/ac/TonellaRM14}. 
%
The manual development of E2E test automation scripts is notoriously costly and time-consuming in practice~\cite{Christophe2014}. To address this, researchers have proposed automated test generation techniques, many of which rely on constructing state-based web application models~\cite{biagiola2020dependency,biagiola2017search,2019-Biagiola-FSE-Dependencies,2019-Biagiola-FSE-Diversity,mesbah:tse12,mesbah:tweb12}. These model-based web testing approaches systematically explore a web application using a crawler to infer a model that captures its functional behavior in terms of states (logical web pages, representing different functionalities) and transitions between them, triggered by user interactions such as clicks~\cite{leithner2020xiev,mesbah:tweb12}. 

Effective models should be both \textit{complete} (i.e., capture all distinct logical pages, adequately covering web app functionality) and \textit{concise} (i.e., avoid redundant representations of the same states)~\cite{2020-Yandrapally-ICSE,Yandrapally-TSE}. In practice, however, models obtained via crawling are often affected by \textit{near-duplicate} states, i.e., replicas of the same logical page differing only by minor changes \cite{2020-Yandrapally-ICSE}.

Near-duplicates negatively affect the model's conciseness, because the same logical state is represented multiple times, and completeness, because the crawler, operating under a finite exploration budget, spends resources revisiting redundant states instead of discovering new ones. This has negative effects on test generation: test suites derived from such models tend to be less effective~\cite{2020-Yandrapally-ICSE, Yandrapally-TSE}. Undiscovered states remain untested, reducing adequacy measures such as code coverage, while redundant test cases targeting duplicate states offer little to no contribution in expanding behavioral coverage~\cite{2019-Biagiola-FSE-Dependencies}.

\begin{figure*}[t]
    \includegraphics[trim={2cm 2cm 1cm 1cm}, clip, width=\linewidth]{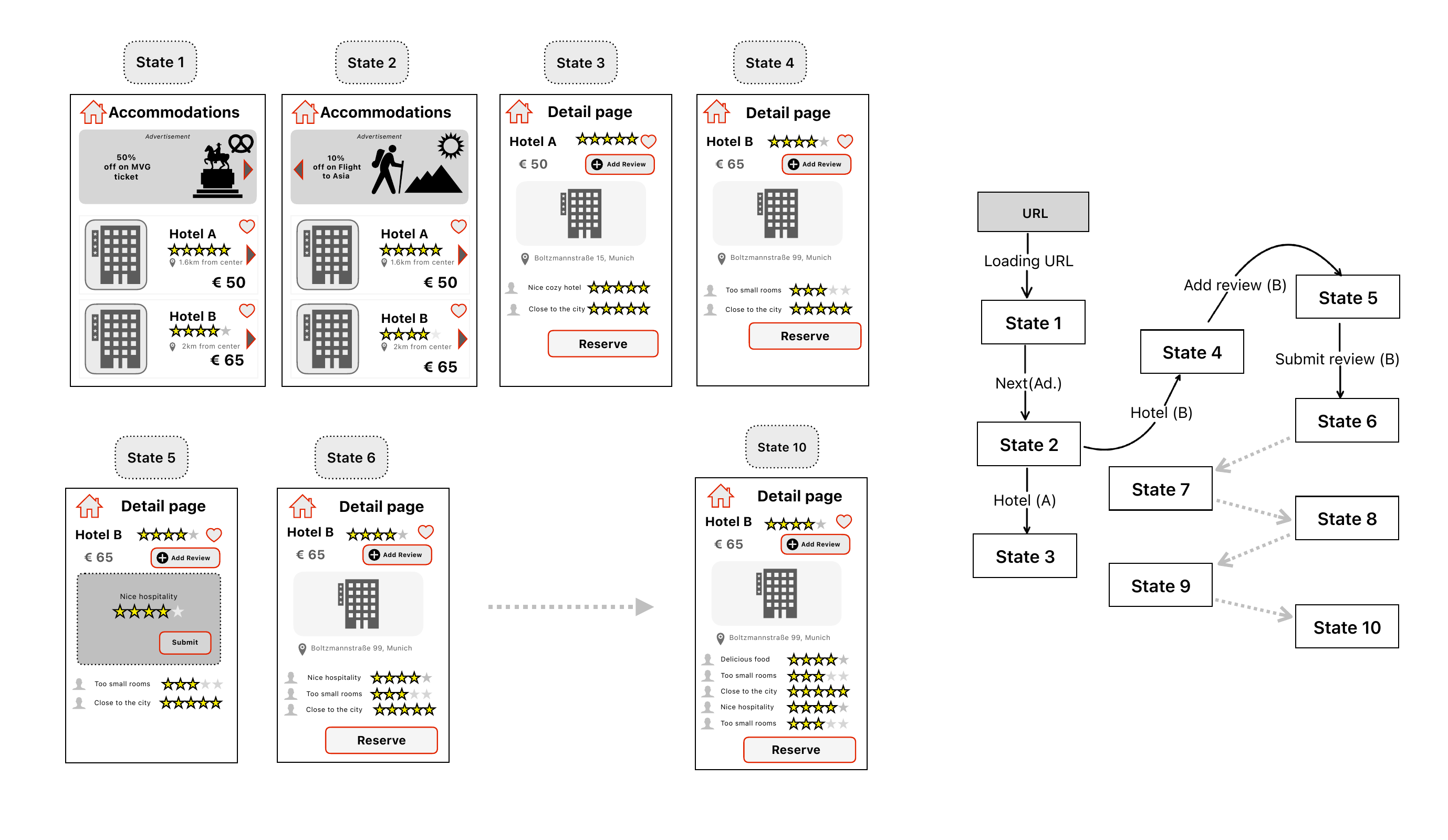}
    \caption{Left: Hotel reservation web app: app states with actionable outlined. Right: An inferred model with near-duplicates.}
    \label{fig:near-duplicates-example}
\end{figure*}

To reduce near-duplicate states, model-based testing relies on a state abstraction function (SAF) to identify and discard redundant states during crawling. Existing SAFs use Document Object Model (DOM), visual, or hybrid similarity metrics, but studies show their effectiveness is highly application-dependent, making it challenging to design a universally reliable SAF~\cite{2020-Yandrapally-ICSE}.

This paper introduces \tool, a novel SAF that combines neural embeddings with transformer-based Siamese neural networks (SNNs). Our approach encodes each web page into an $n$-dimensional embedding derived from token sequences of its markup and textual content. These embeddings are used to train SNN classifiers with distance-based losses (e.g., contrastive or triplet loss), enabling the model to learn semantic similarity within a shared embedding space.
Unlike traditional classifiers, SNNs learn a general similarity function rather than relying on fixed class boundaries, allowing them to generalize to unseen states, an important property for web model inference, where labeled data and balanced class distributions are often unavailable. During crawling, each newly encountered page is embedded and compared to existing states; the classifier then maps its similarity score to near-duplicate or distinct, ensuring that only new states are retained in the inferred model.

We empirically evaluate \tool using established web application benchmarks covering a diverse set of domains. Three tasks are considered: near-duplicate detection, model coverage, and test generation. Specifically, we assess four embedding models—Doc2Vec~\cite{DBLP:journals/corr/LauB16}, BERT~\cite{devlin2019bertpretrainingdeepbidirectional}, ModernBERT~\cite{warner2024smarterbetterfasterlonger}, and MarkupLM~\cite{li2022markuplmpretrainingtextmarkup}—and train two SNN variants per model using Binary Cross-Entropy (BCE)~\cite{jm3} and Triplet Loss~\cite{triplet_1, triplet_2}. 
Across all tasks, SNNs trained with Doc2Vec embeddings and BCE loss consistently achieve the best performance, with statistically significant improvements over four state-of-the-art SAFs. Notably, this configuration attains an $F_1$ score of 96\% in near-duplicate detection (a 45\% improvement) and 81\% in model coverage (a 57\% improvement). Furthermore, tests generated from SNN-based SAF models yield up to 21\% higher code coverage across applications, with an average increase of 12\% compared to current leading techniques.


Our paper makes the following contributions:

\begin{description}[noitemsep]
\item [Technique.] A novel approach called \tool that uses neural embeddings and Siamese neural networks for web model inference and testing, which is available~\cite{replication-package}.
\item [Evaluation.] An empirical study shows that \tool is more effective than three state-of-the-art SAFs in the near-duplicate detection, model coverage, and test generation tasks under different configurations.
\end{description}

\section{Background}\label{sec:background}

\subsection{Automated Web Model Inference}\label{sec:background:modelinference}

Automated web model inference techniques, such as crawling, construct structural models of web applications by systematically exploring their GUI states. This process involves triggering user-driven events (e.g., clicks) and generating input values that cause the application to transition between different pages or views. When a significant change in the rendered content is observed, a new abstract state is added to the inferred model. Each state serves as a generalization of all dynamic runtime instances of a logical page, typically characterized by their underlying DOMs. The resulting model comprises a set of abstract states representing logical web pages and a set of transitions (edges) that capture possible navigations or interactions connecting them.

Consider a simple hotel reservation web application that displays available accommodations for the reservation. Users can view hotel details, add reviews, and make reservations (\autoref{fig:near-duplicates-example}, left). They can navigate from the Listings page to the Details page by clicking on a hotel item. On the Details page, users have the option to write a review or proceed with making a reservation. Additionally, users can return to the Details page from any other page by clicking the Home icon. After exploring these states, the model is generated by constructing a graph that records the reached states as nodes and the transitions as the edges of the graph (\autoref{fig:near-duplicates-example}, right).

\subsection{Near-Duplicate States}

During exploration, newly encountered states fall into three categories. Distinct states exhibit at least one functional or semantic difference from previously explored states (e.g., State 2 vs. State 3) and must be added to the model. Clone states are exact replicas with no functional or semantic differences and are therefore discarded. The most challenging category, Near-duplicate states, are functionally similar pages whose minor variations do not affect overall behavior~\cite{Yandrapally-TSE}. Identifying such states is essential, as they add limited testing value. Indeed, 
web crawlers can optimize their exploration process by skipping near-duplicate states and prioritizing diverse states, leading to inferred models that are more representative and less redundant. 

Existing work on near-duplicate detection~\cite{2020-Yandrapally-ICSE} distinguishes three classes: \emph{Cosmetic} (\(\text{ND}_{1}\)), involving purely visual changes that do not affect functionality; \emph{Dynamic Data} (\(\text{ND}_{2}\)), where pages share the same template but differ in dynamically populated content; and \emph{Structural} (\(\text{ND}_{3}\)), where one page contains additional elements whose functionality and semantics are fully represented in the other.

Prior approaches for near-duplicate detection rely on structural, visual, or hybrid heuristics that require handcrafted rules or similarity threshold tuning.
In this paper, we present a new approach that learns semantic similarity through neural embeddings and SNNs, avoiding the need for manual calibration and thereby enabling more scalable integration in the automatic web test generation process.

\section{Embedding Models} \label{sec:embedding-models}

In this work, we evaluate several embedding methods, ranging from static representations (Doc2Vec) to transformer-based, context-aware models (BERT, ModernBERT, and MarkupLM), as they are suited to capture both semantic and structural dependencies within web pages, key aspects to distinguishing distinct, near-duplicate, and clone states. The main characteristics of each method are summarized below.

\subsection{Doc2Vec}

Doc2Vec~\cite{DBLP:journals/corr/LauB16} is a static embedding technique that generates a single dense vector for an entire document without explicit token-level context. Unlike BERT-based models, Doc2Vec does not require subword or byte-pair tokenization. Instead, it processes the entire document as a sequence of words and learns a fixed-length representation for each document. This approach is typically more efficient for large datasets, but the downside is that it may lack the nuanced context-dependent representation that BERT-like models provide. 

\subsection{BERT}

BERT~\cite{devlin2019bertpretrainingdeepbidirectional} relies on the WordPiece~\cite{song2021fastwordpiecetokenization} subword tokenization algorithm to split input text into smaller units, limiting each sequence to a maximum length of 512 tokens. Any token exceeding this limit is truncated by the chunking process detailed in \autoref{sec:chunking}. After tokenization, the tokens are passed to the BERT model to generate embeddings, and this process is repeated for all chunks. The final input embedding is constructed by averaging the embeddings of all chunks into a single vector. We use the averaging technique for BERT embeddings because the dimensionality increases as the input size grows, which would cause our model to become larger. 





\subsection{ModernBERT}

ModernBERT~\cite{warner2024smarterbetterfasterlonger} also relies on the WordPiece~\cite{song2021fastwordpiecetokenization} subword tokenization algorithm to split input text into smaller units, but with a larger context, as it can process sequences of up to 8,192 tokens. This extended context window is more than sufficient for the data considered in this work, and no truncation is therefore needed.

\subsection{MarkupLM}

MarkupLM~\cite{li2022markuplmpretrainingtextmarkup} is a transformer model tailored for HTML documents. Building upon BERT, it encodes both textual and structural information by incorporating markup-aware features such as tag hierarchies, DOM positions, and XPath representations. Unlike standard BERT tokenizers that treat tags as plain text, MarkupLM preserves HTML structure during tokenization, allowing it to capture relationships between nodes and their content.
Similar to BERT, inputs are limited to 512 tokens, but MarkupLM's HTML-aware positional encoding enables the model to represent hierarchical layouts that standard transformers can not~\cite {NEURIPS2019_6e091746}. In our setup, each node's text and XPath are jointly embedded to form structure-aware vector representations. These embeddings, concatenated across document chunks, yield richer, context-sensitive encodings of web pages for downstream similarity and modeling tasks.

\section{Siamese Neural Networks}\label{sec:SNN}

Siamese Neural Networks (SNNs) are neural architectures designed to learn similarity relationships between pairs of inputs by projecting them into a shared embedding space. Originally developed for tasks such as signature and face verification, SNNs have shown strong performance in domains requiring precise discrimination between similar but non-identical instances. Their strength lies in learning a generalized similarity function rather than fixed class boundaries, making them particularly suitable for web application state comparison, where new, previously unseen states frequently emerge, and labeling all possible variations is infeasible.

We train two variants of SNN, each adopting a different loss function: one using Binary Cross-Entropy (BCE)~\cite{jm3} and the other using Triplet Loss~\cite{triplet_1, triplet_2}. 
In the following, we detail these two training strategies and provide the mathematical formulations for each loss function.

\subsubsection{BCE-Based SNN}
\label{sec:bce_siamese}

\begin{figure}[htpb]
  \centering
  \includegraphics[width=\linewidth]{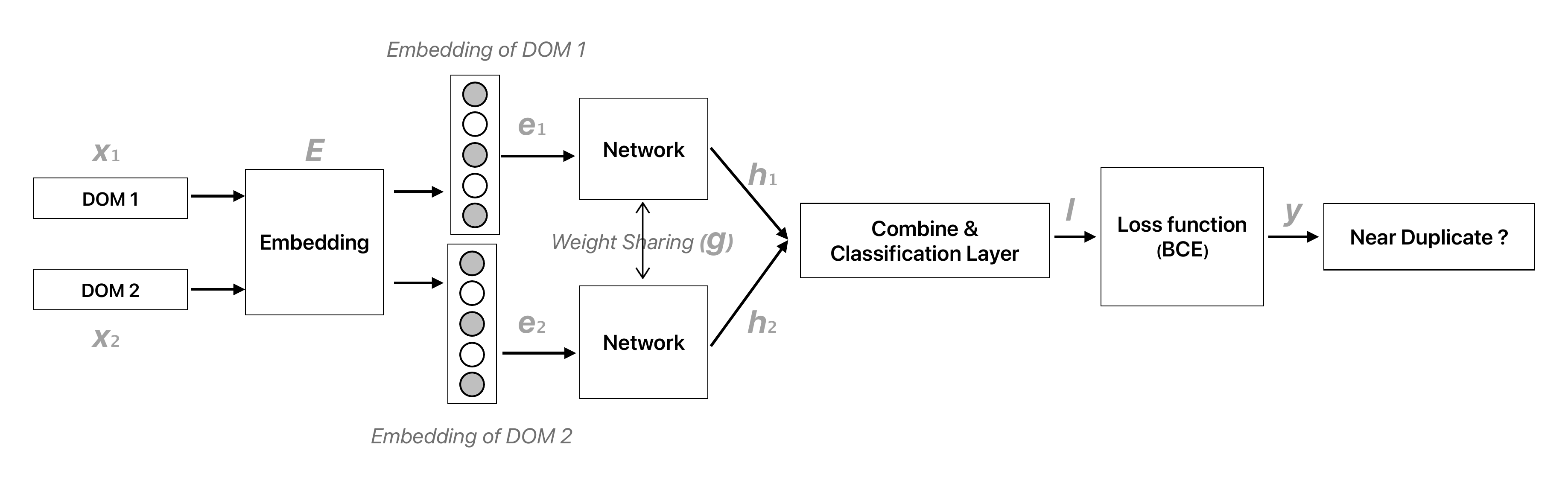}
  \caption[BCE Network]{BCE-based SNN Architecture.}
  \label{fig:net1}
\end{figure}

The BCE-based SNN architecture is displayed in \autoref{fig:net1}. Initially, a pair of input samples, $(x_1, x_2)$, is transformed into initial embeddings using a selected embedding technique \( E \). These embeddings, namely $\mathbf{e}_1 = E(x_1)$ and $\mathbf{e}_2 = E(x_2)$, are then passed through a \emph{shared} feature extraction network \( g \), yielding two lower-dimensional representations, $\mathbf{h}_1 = g(\mathbf{e}_1)$ and $\mathbf{h}_2 = g(\mathbf{e}_2)$, where \( g \) denotes the learned task-specific embedding function, typically implemented as a feed-forward neural network. To explicitly model the relationship between the embeddings, these hidden representations are combined into a single representation \(\mathbf{h}_{\text{combined}}\), as follows:

\[
\mathbf{h}_{\text{combined}} = [\,\mathbf{h}_1; \mathbf{h}_2; |\mathbf{h}_1 - \mathbf{h}_2|; (\mathbf{h}_1 \odot \mathbf{h}_2); \text{sim}(\mathbf{h}_1, \mathbf{h}_2)\,],
\]

\noindent
where \([\,\cdot\,;\,\cdot\,]\) denotes concatenation, \(\odot\) is the element-wise product, and \(\text{sim}(\mathbf{h}_1, \mathbf{h}_2)\) is the cosine similarity metric. The combined representation \(\mathbf{h}_{\text{combined}}\) is then fed into a small classification layer \( c \), which computes a logit $\ell = c(\mathbf{h}_{\text{combined}}) = W\mathbf{h}_{\text{combined}} + b$, where \( W \) and \( b \) represent the learned weights and biases in the classification layer, respectively. Finally, the logit \(\ell\) is passed through a sigmoid activation function \(\sigma\), producing the estimated probability \(\hat{y} = \sigma(\ell) \) that the input pair represents a near-duplicate pair.
Let \( y \in \{0,1\} \) be the ground-truth label, where \(y=1\) indicates a near-duplicate pair, and \(y=0\) indicates a distinct pair. We optimize the network parameters by minimizing the~BCE loss:
\[
\mathcal{L}_{\text{BCE}} = - \frac{1}{N}\sum_{i=1}^{N} \bigl[\,y_i \,\log\!\bigl(\sigma(\ell_i)\bigr) \;+\; (1 - y_i)\,\log\!\bigl(1 - \sigma(\ell_i)\bigr)\bigr],
\]
where \(N\) is the total number of training pairs, \(\ell_i\) is the predicted logit for the \(i\)-th pair, and \(\sigma(\ell_i)\) is the output probability after applying the sigmoid function. By minimizing this loss, the network learns to assign higher probabilities to near-duplicate pairs and lower probabilities to distinct pairs.

\subsubsection{Triplet-Loss-Based SNN}
\label{sec:triplet_siamese}

\begin{figure}[htpb]
  \centering
  \includegraphics[width=\linewidth]{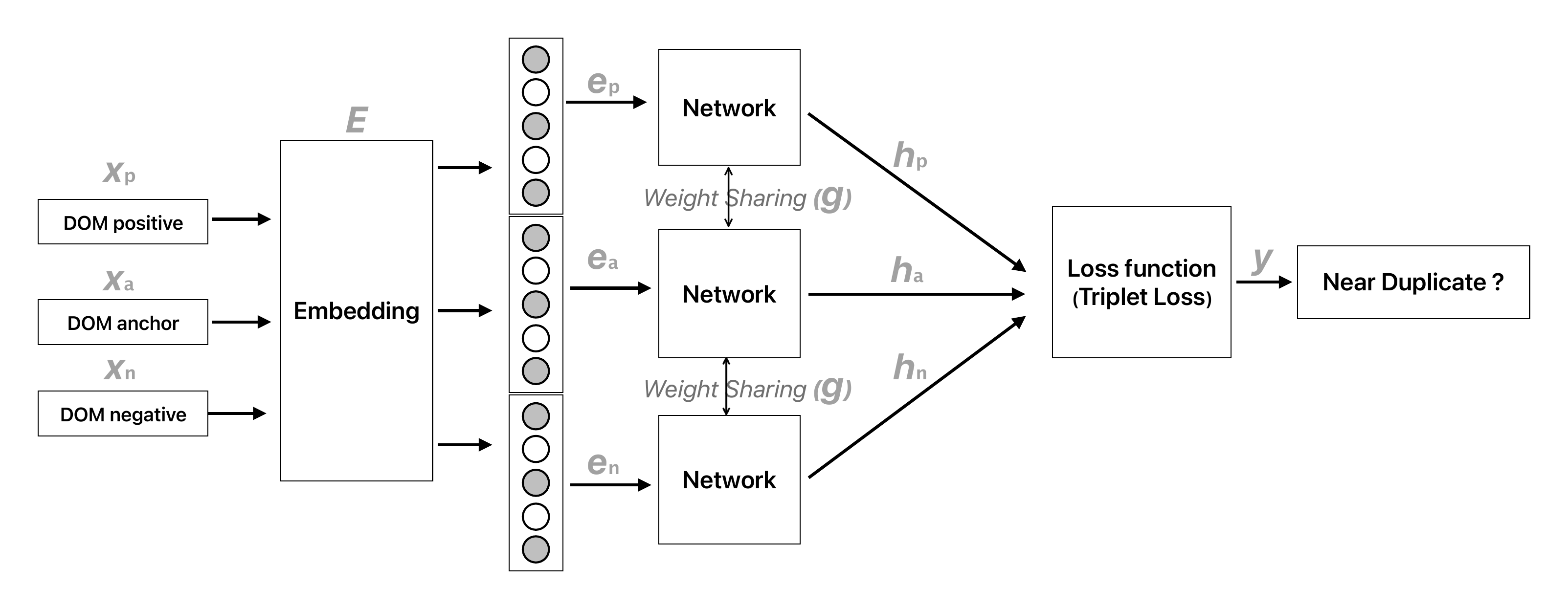}
  \caption[Triplets Network]{Triplet-Based~SNN Architecture}
  \label{fig:net2}
\end{figure}

\begin{figure*}[t]
\includegraphics[trim={0cm 0cm 0cm 0cm}, clip, width=\linewidth]{_figures/snn_overview.pdf}
\caption{Overview of our approach.}
\label{fig:dom2vec_approach}
\end{figure*}

In the Triplet-Based SNN approach as visible in \autoref{fig:net2}, we enforce \emph{relative distance} constraints rather than directly predicting a binary label. Each training sample is organized into a triplet \(\bigl(x_a, x_p, x_n\bigr)\), where \(x_a\) is an \textit{anchorDOM}, \(x_p\) is a \textit{positiveDOM}, and \(x_n\) is a \textit{negativeDOM}. A shared embedding function \( g \) projects these inputs into a latent space, i.e., $\mathbf{h}_a = g(E(x_a))$, $\mathbf{h}_p = g(E(x_p))$, and $\mathbf{h}_n = g(E(x_n))$.
We then compute the squared Euclidean distances $D_{ap} = \left\|\mathbf{h_a} - \mathbf{h_p}\right\|_2^2$, $D_{an} = \left\|\mathbf{h_a} - \mathbf{h_n}\right\|_2^2$.
The triplet loss encourages \(\mathbf{a}\) to be closer to \(\mathbf{p}\) than to \(\mathbf{n}\) by at least a margin \(\alpha\). Formally, the triplet loss is given by:
\[
\mathcal{L}_{\text{triplet}} = \frac{1}{M} \sum_{i=1}^{M} \max\bigl(0,\, \alpha + D_{ap}^i - D_{an}^i\bigr),
\]
where \(M\) is the number of triplets and \(\alpha\) is a hyperparameter that controls how strongly the positive distance must be smaller than the negative distance. This objective naturally clusters near-duplicate samples in the embedding space and pushes distinct samples farther apart, without requiring an explicit probability output. We set \(\alpha\) to \(1.0\), the default parameter provided in the original paper.

During inference, we use the shared weights learned during training to compute embeddings directly for any two given samples (without requiring a triplet). We then measure their distance in the embedding space and compare this distance to a standard threshold \(0.5\), selected through pilot experiments. If the distance is below this threshold, the pair is classified as a near-duplicate; otherwise, it is considered distinct. 

\section{Approach}\label{sec:approach}


\autoref{fig:dom2vec_approach} illustrates our approach, which consists of four phases, namely (1)~Embedding Pipeline, (2)~Dataset Creation and Training Pipeline, (3)~Crawling, and (4)~Test Creation. 

%
In the Embedding Pipeline phase, we start from a labeled corpus of web pages where pairs are annotated as either distinct or clone/near-duplicate. For each distinct web page (GUI state), we extract a token-sequence representation of the DOM -- a combined representation of \textit{content+tags} -- and then compute embeddings using a pre-trained model.
After computing the embedding representations of the distinct web pages, we use them in the subsequent Dataset Creation and Training Pipeline phase to train SNN classifiers to distinguish distinct and clone/near-duplicate web pages.
In the third phase, Crawling, the trained SNN-based classifiers are deployed as runtime state abstraction functions (SAFs). For each newly encountered web page, our approach computes its embedding, compares it against embeddings of existing states in the model, and uses the classifier to predict whether the page represents a distinct state or a near-duplicate of an existing one. 
%
In the fourth phase, Test Generation, each crawl path in the model inferred in the previous phase is turned into a web test case. We now detail each phase of our approach.

\subsection{Embedding Pipeline}

\subsubsection{Data Preprocessing}

Our approach requires computing embeddings for
web pages. 
We adopt a combined representation based on \textit{content+tags}, as it retains both textual and structural information for web page understanding. 
We apply an initial set of preprocessing steps to all embedding models considered in this work, namely: Doc2Vec~\cite{DBLP:journals/corr/LauB16}, BERT~\cite{devlin2019bertpretrainingdeepbidirectional}, ModernBERT~\cite{warner2024smarterbetterfasterlonger} and MarkupLM~\cite{li2022markuplmpretrainingtextmarkup}. 
First, we remove any \texttt{<script>} and \texttt{<style>} tags, HTML comments, and HTML attributes that do not contribute to a page's primary content or structure. 


\begin{lstlisting}[caption={HTML of the My Reservation page (State 7)}, label={lst:state7-html}]
<html lang="en">
<head>
  <title>Reservations</title>
  <link rel="stylesheet" href="styles.css"></link>
  <script src="utils.js"></script>
</head>
<body>
  <div class="reservations-container">
    <h2 class="header">
      <i class="home-icon">&#8962;</i>
      My Reservations
    </h2>
    <div class="reservation-card">  <!-- Reservation 1 -->
      <img src="hotel-icon.png" alt="Hotel" class="hotel-icon">
      <h3 class="hotel-name">Hotel B</h3>
      <p class="stay-details">25 Apr - 30 Apr, Munich</p>
      <span class="status ongoing">On going</span>
      <i class="icon heart-icon">&#9825;</i>
      <i class="icon arrow-icon">&#9654;</i>
    </div>
    <div class="reservation-card">  <!-- Reservation 2 -->
      <img src="hotel-icon.png" alt="Hotel" class="hotel-icon">
      <h3 class="hotel-name">Hotel C</h3>
      <p class="stay-details">1 Jan - 5 Jan, Munich</p>
      <span class="status completed">Completed</span>
      <i class="icon heart-icon">&#9825;</i>
      <i class="icon arrow-icon">&#9654;</i>
    </div>
  </div>
</body>
</html>
\end{lstlisting}

As shown in \autoref{lst:state7-html}, the HTML snippet contains additional \emph{tags} (e.g., \texttt{<script>} and \texttt{<link>}), \emph{comments}, and \emph{attributes} that we remove in the first preprocessing step. We then retain only the \emph{tags} and \emph{text content} and remove special characters such as \texttt{<}, \texttt{!}, \texttt{/}, and \texttt{>}. This final token sequence preserves the essential structural and textual elements needed for our downstream embedding models while excluding \emph{tags}, \emph{attributes}, and content deemed non-essential.
For Doc2Vec, BERT and ModernBERT, the input token representation is:

\noindent
\begin{quote}
\footnotesize
\texttt{[\,html, head, title, Reservations, title, head, body, div, h2, i, \&\#8962;, i, My, Reservations, h2, div, img, h3, Hotel, B, h3, p, 25, Apr, -, 30, Apr, Munich, p, span, On, Going, i, \&\#9825;, i, i, \&\#9654;, i, div, img, h3, Hotel, C, h3, p, 1, Jan, -, 5, Jan, Munich, p, span, Completed, span, i, \&\#9825;, i, i, \&\#9654;, i, div, div, body, html\,]}
\end{quote}

Since  MarkupLM~\cite{li2022markuplmpretrainingtextmarkup} is trained directly on HTML pages rather than plain text, we provide it with each web page's node content as a list alongside their corresponding XPath to obtain embeddings. For the example in \autoref{lst:state7-html}, the MarkupLM input token representation is shown below.

\noindent
\text{Node Sequence:}
\begin{quote}
\footnotesize
\texttt{[Reservations, \&\#8962;, My, Reservations, Hotel, B, 25, Apr, -, 30, Apr, Munich, On, going, \&\#9825;, \&\#9654;, Hotel, C, 1, Jan, -, 5, Jan, Munich, Completed, \&\#9825;, \&\#9654;]}
\end{quote}

\noindent
\text{XPath Sequence:}

\begin{quote}
\footnotesize
\begin{lstlisting}
[/html/head/title, /html/body/div/h2/i, /html/body/div/h2, /html/body/div/h2, /html/body/div/div[1]/h3, /html/body/div/div[1]/h3, /html/body/div/div[1]/p, /html/body/div/div[1]/p, /html/body/div/div[1]/p, /html/body/div/div[1]/p, /html/body/div/div[1]/p, /html/body/div/div[1]/p, /html/body/div/div[1]/span, /html/body/div/div[1]/span, /html/body/div/div[1]/i, /html/body/div/div[1]/i, /html/body/div/div[2]/h3, /html/body/div/div[2]/h3, /html/body/div/div[2]/p, /html/body/div/div[2]/p, /html/body/div/div[2]/p, /html/body/div/div[2]/p, /html/body/div/div[2]/p, /html/body/div/div[2]/p, /html/body/div/div[2]/span, /html/body/div/div[2]/i, /html/body/div/div[2]/i]
\end{lstlisting}
\end{quote}

\subsubsection{Chunking}  \label{sec:chunking}

Doc2Vec embeddings are static and can handle inputs of any length.
ModernBERT can also handle 8,192 sequence lengths, which provides a sufficient context for the data we consider in this work. However, BERT and MarkupLM embeddings are constrained by a maximum input size of 512 tokens. For large documents, such as lengthy HTML files, any token beyond this limit is ignored during embedding generation, leading to potential loss of relevant information. Even after our preprocessing steps, we observe that approximately 34\% of tokens for BERT are trimmed on average, whereas MarkupLM trims about 6\% (\autoref{tab:average_truncated_percentage}).

\begin{table}[b]
  \caption{Average Truncated Percentage.}
  \label{tab:average_truncated_percentage}
  \centering
  \setlength{\tabcolsep}{6.5pt}
  \renewcommand{\arraystretch}{1.0}
  {\footnotesize
  \begin{tabular}{l c c c c}
    \toprule
    \textbf{Embedding} & \textbf{No Chunking} & \textbf{2 Chunks} & \textbf{3 Chunks} & \textbf{4 Chunks} \\
    \midrule
    Bert Base    & 34\% & 18\% & 15\% & 3\% \\
    MarkupLM            & 6\%  & 1\%  & 0\%  & 0\% \\
    \bottomrule
  \end{tabular}
  }
\end{table}

To mitigate this limitation, we introduce chunking. First, we split each document into chunks of up to 512 tokens, with a defined overlap to maintain contextual coherence across chunk boundaries. This overlap helps ensure that important information spanning the boundary between two chunks is not lost. If a document does not reach the maximum number of chunks, we pad the remaining chunks with zeros so that all inputs maintain a consistent dimensionality during training.

\begin{figure}[t]
  \centering
  \includegraphics[trim={1.5cm 1.5cm 1.5cm 1.5cm}, clip,width=.9\linewidth]{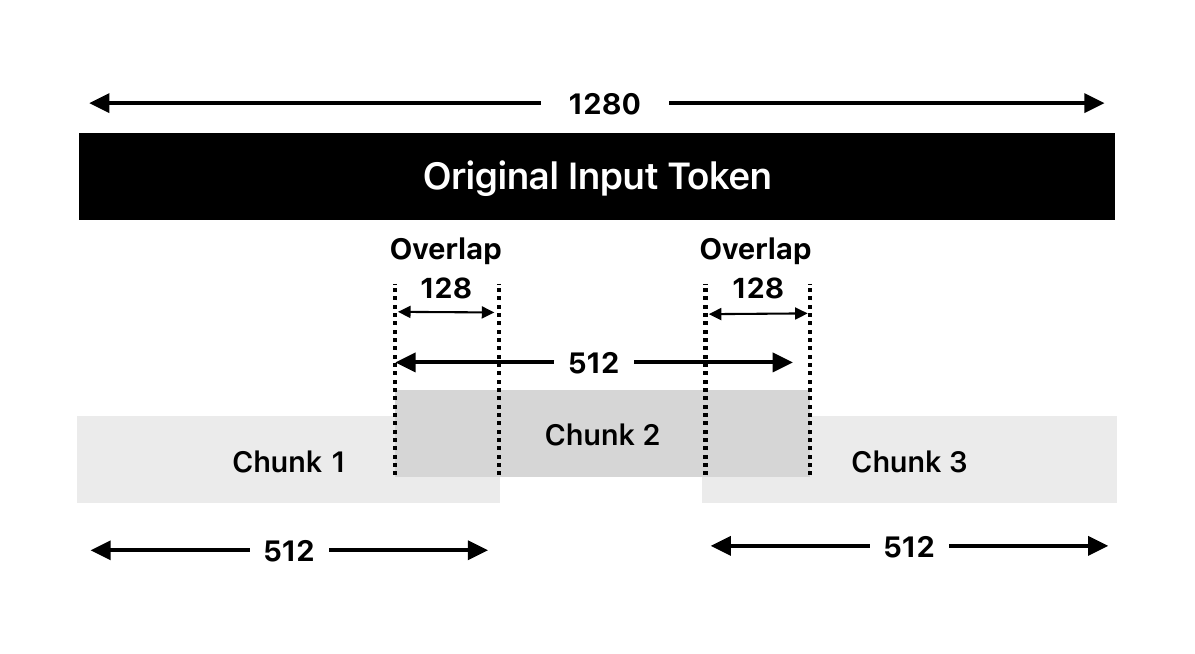}
  \caption[Chunking]{Chunking with a 128-token overlap.}
  \label{fig:chunking}
\end{figure}

Too many chunks can lead to excessive zero padding, which wastes computational resources and may negatively impact model training. To address this, we introduce a maximum number of chunks, i.e., the \emph{chunk threshold}, which is a hyperparameter that we tune empirically. For instance, a sequence of 1280 tokens can be split into three chunks, each with a 128-token overlap, as illustrated in \autoref{fig:chunking}. After applying chunking, the average truncation for ~BERT is reduced from 34\% to 18\% with 2 chunks, and with four chunks, it can be lowered to just 3\%. For MarkupLM, using two chunks decreases its truncation rate from 6\% to a mere 1\%, as indicated in \autoref{tab:average_truncated_percentage}.

\subsubsection{Tokenization and Embedding}
\label{subsec:embedding}

In this work, we perform classification tasks without fine-tuning the transformer-based models to assess whether these rich embeddings are sufficient for our downstream task of web page understanding.

For Doc2Vec, we used a pre-trained embedding model 
to generate embeddings~\cite{web2vec}, as this model is already fine-tuned for generating embeddings for HTML inputs. 

For BERT, ModernBERT, and MarkupLM, we used the given tokenizer (available on \texttt{HuggingFace}) to convert the input into a sequence of tokens.

BERT and MarkupLM limit each sequence to a maximum length of 512 tokens. We addressed longer sequences using chunking (\autoref{sec:chunking}). After tokenization, the tokens are passed to the corresponding models to generate embeddings, and this process is repeated for all chunks. The final input embedding is constructed by averaging the embeddings of all chunks into a single vector. 




On the other hand, ModernBERT can process sequences of up to 8,192 tokens. This extended context window is more than sufficient for the data considered in this work; hence, we do not employ chunking.

\begin{algorithm}[t]
  \scriptsize
  \SetKwProg{Fn}{Function}{:}{}
  \SetKw{Break}{break}
  \SetKwComment{Comment}{$\triangleright$\ }{}
  
  {\textsc{extractTokens}({$n$})}{
    let $\mathit{tokens}$ be an empty list\; 
    $\mathit{tokens}$.$\mathit{append}(\mathit{getTokens}(n))$\;
    \ForEach(){children node $c$ of $n$, from left to right}{
      \If(){$c$ is not a script, style, or comment node}{
        $\mathit{tokens}.\mathit{append}(\textsc{extractTokens}(c))$
      }
    }
    \Return{$\mathit{tokens}$}\;
  }
    \BlankLine
  \Fn{\textsc{Crawl}$($initial URL$)$}{
    $s_1 \gets getState($\textit{initial URL}$)$\;
    $\mathit{model} \gets \mathit{initializeModel}(s_1)$\;
    \While(){$\neg$ timeout}{
      $next\gets \mathit{nextStateToExplore}(\mathit{model})$\;
      \If(\Comment*[f]{app exhaustively explored}){$\mathit{next}=\mathit{nil}$}{
        \Break\;
      }
      $s\gets \mathit{getToState}(\mathit{next})$\; 
      \For(){$e\in \mathit{getCandidateEvents}(s)$}{
        $fireEvent(e)$\;
        $s_c\gets$ current state after firing the event $e$\;
        \If(){$\neg~\textsc{IsDuplicate}(s, \mathit{model})$}{
          add $s_c$ to $\mathit{model}$\; 
        }
      }
    }
    \Return{$model$}\;
  }
    \BlankLine
    \Fn{\textsc{IsDuplicate}({$s_c$, $\mathit{model}$})}{
    \ForEach(){state $s^\prime$ in $\mathit{model}$}{
      \If(){\textsc{Classify}($s_c$, $s^\prime$) = `\textit{clone}'}{
        \Return{True}\Comment*[r]{$s$ is a duplicate of $s^\prime$}
      }
    }
    \Return{False}\;    
  }
    \BlankLine
  \Fn(\Comment*[f]{$EM$: embedding model}){\textsc{Classify}({$p_1$, $p_2$, $EM$})}{
    $r_{1}\gets \textsc{extractTokens}(p_1.\mathit{getRootNode}())$\;
    $r_{2}\gets \textsc{extractTokens}(p_2.\mathit{getRootNode}())$\;
    $e_{1}\gets EM.computeEmbedding(r_{1})$ \;
    $e_{2}\gets EM.computeEmbedding(r_{2})$\; 
    $s \gets \mathit{cosineSimilarity}(e_1, e_2)$\;
    
    \Return{$\mathit{classifier}.\mathit{classify}(s)$}\;
  }
  \caption{Web App Crawling with SNNs}\label{algo:crawling}
\end{algorithm}



\subsection{Training State Abstraction Function}

In the second phase, we train the SAF using a labeled corpus of web page pairs, where each pair is annotated to indicate whether the pages are clones or near-duplicates.
For each pair, we generate embeddings using one of the selected embedding models and compute their cosine similarity~\cite{journals/debu/Singhal01}, a standard metric for measuring vector similarity. These embeddings and similarity scores are then used to train an SNN-based classifier to distinguish between near-duplicate and distinct web pages. For simplicity, clones are grouped under the near-duplicate category, as their detection does not require additional sophistication beyond our similarity analysis.


\subsection{Crawling and Model Inference}
The third phase consists of using the trained SAF during crawling to infer crawl models that can be used for automated test generation. 

\subsubsection{The Crawler}

The crawler loads the web pages in a web browser and exercises client-side JavaScript code to simulate user-like interactions with the web app. This allows the crawler to support modern, client-side intensive, single-page web applications.
The main conceptual steps performed when exploring a web application are outlined in the \textsc{Crawl} function of \autoref{algo:crawling} (lines~8---21).

Crawling starts at an initial URL, the homepage is loaded into the browser and the initial DOM state, called index, is added to the model (line~9). Subsequently, the main loop (lines~11---20) is executed until the given time budget expires or there are no more states to visit (i.e., the web app has been exhaustively explored according to the crawler). 
In each iteration of the main loop, the first unvisited state in the model is selected (line~12), and the crawler puts in place adequate actions to reach said state. If the state cannot be reached directly, it retrieves the path from the index page and fires the events corresponding to each transition in the path. Upon reaching the unvisited state, the clickable web elements are collected (i.e., the web elements on which interaction is possible, line~16), and user events such as filling forms or clicking items are generated (line~17). After firing an event, the current DOM state $s_c$ is captured (line~18).
The \textsc{isDuplicate} function supervises the construction of the model and checks whether $s_c$ is a duplicate of an existing state (lines~22---26) by computing pairwise comparisons with all existing states in the model using the SAF. The state $s_c$ is eventually added to the model if the SAF regards it as a distinct state, i.e., a state that is not a duplicate of another existing state in the model (lines~23---26). Otherwise, it is rejected, and the crawler continues its exploration from the next available unvisited state until the timeout is reached. 

\subsubsection{Usage of the State Abstraction Function}

The \textsc{Classify} procedure (lines~27---33) illustrates the SAF. Given two web pages $p_1$, $p_2$, we first extract the token-sequence representation from each page based on the selected embedding model $EM$, obtaining the list of tokens for each web page (lines~30---31). 
Each of the two token sequences $r_{1}$ and $r_{2}$ is then fed to the appropriate embedding model  (line 32) to compute an embedding (lines~33---34). Then, the cosine similarity between the two resulting embeddings $e_{1}$ and $e_{2}$ is computed, obtaining a similarity score that is appended to the list $s$ of similarities computed so far (line~35). 
Next, the classifier marks the two pages as either distinct or clones based on the list $s$ of similarity scores, and determines the SAF return value (line~33), which is `\textit{clone}' in case of near-duplicate detection or `\textit{distinct}' otherwise. 

\head{Example}
Consider the following embeddings produced for our running example, i.e., embedding on content+tags and $EM$ = [`\textit{BERT}']:
{
\begin{align} 
p_1 &= \textrm{Accomodations Page}   && e_1 = [-0.45, 0.56, ..., 0.30] \nonumber \\ 
p_2 &= \textrm{Detail Page Hotel A}  && e_2 = [-0.55, 0.17, ..., 0.90] \nonumber \\
p_3 &= \textrm{Detail Page Hotel B} && e_3 = [-0.56, 0.19, ..., 0.95] \nonumber
\end{align}
}%

During crawling, let us assume that a decision tree classifier flags a pair of pages as `\textit{near-duplicate}' when the cosine similarity between their embeddings satisfies the root decision node condition ($s > 0.8$). If $sim(e_1, e_2) = 0.56$,  $p_2$ is added to the model, as $p_2$ is not too similar to $p_1$. Then, when exploring $p_3$, we obtain $sim(e_3, e_1) = 0.58$ and $sim(e_2, e_3) = 0.95$. Hence,   page $p_3$ is not added to the model as it is recognized as a near-duplicate of $p_2$. 

\subsection{Test Creation}

Our approach automatically generates a test suite during crawling through \textit{segmentation}~\cite{biagiola2020dependency}. 
The crawl sequence of states is segmented into test cases when (1)~the current DOM state no longer contains any candidate clickable elements to be fired and the crawler is reset to the index page; 
(2)~no new states are present on the current path. 
In the case of \autoref{fig:near-duplicates-example} (right), four (redundant) test cases are generated, one for each state representing the Detail page for item A. With our SAF, the output model only has one state for the Detail page. Hence, only one test would be generated, reducing redundancy while keeping model and code coverage the same. 

%





\section{Empirical Study}\label{sec:study}

\subsection{Research Questions}

To assess the practical benefits of \tool for web testing, we consider the following research questions. 

\noindent
\head{RQ\textsubscript{1} (near-duplicate detection)}
\textit{How effective is \tool in distinguishing near-duplicate from distinct web app states?}

\noindent
\head{RQ\textsubscript{2} (model \changed{quality})}
\textit{How do the web app models generated by \tool compare to a ground-truth model?}

\noindent
\head{RQ\textsubscript{3} (code coverage)}
\textit{What is the code coverage of the tests generated from \tool's web app models?}

\noindent
\head{RQ\textsubscript{4} (time efficiency)}
\textit{How efficient is \tool in terms of training and inference time, and its applicability in real-world web crawling?}

RQ\textsubscript{1} aims to assess what configuration (Embedding model and network variant) of our approach, in terms of embedding and classifier, is more effective at detecting near-duplicates through state-pair classification. 
RQ\textsubscript{2} focuses on the crawl model quality in terms of completeness and conciseness. 
RQ\textsubscript{3} evaluates our approach when used for web testing, specifically assessing the test suites generated from the crawl models in terms of code coverage of the web apps under test.
RQ\textsubscript{4} addresses the computational efficiency of our approach. Two sub-questions related to time efficiency are evaluated: the model training duration (RQ\textsubscript{4.1}) and inference time (RQ\textsubscript{4.2}) from 1,000 randomly selected state sample pairs.

\subsection{Datasets}\label{sec:dataset}





\begin{table}[t]
\caption{Subject Set with manual classification.}
\label{table:table-subjects}

\setlength{\tabcolsep}{11.2pt}
\renewcommand{\arraystretch}{1.0}

\begin{tabular}{@{}lrrrrrr@{}}

\toprule

& \rot{\textbf{Logical States}}
& \rot{\textbf{Concrete States}}
& \rot{\textbf{Redundancy (\%)}} 
& \rot{\textbf{State-pairs}} 
& \rot{\textbf{Distinct}} 
& \rot{\makecell[l]{\textbf{Clones and}\\\textbf{Near-duplicates}}} \\ 

\midrule

App\textsubscript{1} & 25 & 131 & 524 & 8,515 & 6,142 & 2,373 \\
App\textsubscript{2} & 36 & 189 & 525 & 17,766 & 14,988 & 2,778 \\
App\textsubscript{3} & 23 & 99 & 430 & 4,851 & 432 & 531 \\
App\textsubscript{4} & 14 & 151 & 1,079 & 11,325 & 7,254 & 4,071 \\
App\textsubscript{5} & 53 & 151 & 285 & 11,325 & 10,206 & 119 \\ 
App\textsubscript{6} & 21 & 153 & 729 & 11,628 & 10,683 & 945 \\
App\textsubscript{7} & 20 & 140 & 700 & 9,730 & 5,782 & 3,948 \\
App\textsubscript{8} & 10 & 150 & 1,500 & 11,175 & 6,569 & 4,606 \\
App\textsubscript{9} & 14 & 149 & 1,064 & 11,175 & 9,411 & 1,615 \\

\bottomrule

\end{tabular}
\end{table}

We used an existing dataset ($\mathcal{SS}$) available from the study by Yandrapally et al.~\cite{2020-Yandrapally-ICSE}.
It contains 97,500 state-pairs of nine subject apps (\autoref{table:table-subjects}), which were also manually labeled by the authors of the study as clone, near-duplicate or distinct. 
These nine web apps (\autoref{table:table-subjects}) have been used as subjects in previous research on web testing~\cite{2016-Stocco-ICWE,2017-Stocco-SQJ,2019-Biagiola-FSE-Diversity,2019-Biagiola-FSE-Dependencies}. Despite being developed with different frameworks, they all provide CRUD functionalities (e.g., login, or add user) which make them functionally similar. Five apps are open-source PHP-based applications, namely 
Addressbook (App\textsubscript{1}, \textit{v. 8.2.5})~\cite{addressbook}, 
Claroline (App\textsubscript{2}, \textit{v. 1.11.10})~\cite{claroline}, 
PPMA (App\textsubscript{3}, \textit{v. 0.6.0})~\cite{ppma}, 
MRBS (App\textsubscript{4}, \textit{v. 1.4.9})~\cite{mrbs} and MantisBT (App\textsubscript{5}, \textit{v. 1.1.8})~\cite{mantisbt}. Four are JavaScript single-page applications--Dimeshift~(App\textsubscript{6}, \textit{commit 261166d})~\cite{dimeshift}, 
Pagekit~(App\textsubscript{7}, \textit{v. 1.0.16})~\cite{pagekit}, 
Phoenix~(App\textsubscript{8}, \textit{v. 1.1.0})~\cite{phoenix} and PetClinic~(App\textsubscript{9}, \textit{commit 6010d5})~\cite{petclinic}--developed using popular JavaScript frameworks such as \textit{Backbone.js}, \textit{Vue.js}, \textit{Phoenix/React} and \textit{AngularJS}.

\subsection{Baselines}

Based on the study by Yandrapally et al.~\cite{2020-Yandrapally-ICSE, fraggen2022}, 
we selected four algorithms as baselines for our approach, one structural, one visual, and one hybrid method. 
The structural algorithm is RTED (Robust Tree Edit Distance)~\cite{PawlikA15}, a DOM tree edit distance algorithm. The visual algorithm is PDiff~\cite{Yee:2001:SSV:383745.383748}, which compares two web page screenshots based on a human-like concept of similarity that uses spatial, luminance, and color sensitivity. The hybrid method is FragGen~\cite{fraggen2022}, a fragment-based approach for near-duplicate detection. FragGen uses a more granular representation of a web page by incorporating both structural and visual aspects. Specifically, it employs a deterministic recursive method built upon the RTED algorithm for structural comparison, and the Visual Histogram algorithm~\cite{histogram} for visual comparison. 

We chose them as baselines for our approach because they are the best structural, visual, and hybrid algorithms for near-duplicate detection~\cite{2020-Yandrapally-ICSE, fraggen2022}.

\subsection{Configurations}\label{sec:configurations}

\begin{table}[t]
    \caption{\tool's configurations.}
    \setlength{\tabcolsep}{15.5pt}
    \renewcommand{\arraystretch}{1.0}
    \label{table:eval-techniques}
  \centering
  \begin{tabular}{l c c}
    \toprule
    \textbf{Name} & \textbf{Embedding Model} & \textbf{SNN Variant/Loss} \\
    \midrule
    SNN (T\textsubscript{1}) & Doc2Vec & BCE \\
    SNN (T\textsubscript{2}) & Doc2Vec & Triplet \\
    SNN (T\textsubscript{3}) & BERT & BCE \\
    SNN (T\textsubscript{4}) & BERT & Triplet \\
    SNN (T\textsubscript{5}) & ModernBERT & BCE \\
    SNN (T\textsubscript{6}) & ModernBERT & Triplet \\
    SNN (T\textsubscript{7}) & MarkupLM & BCE \\
    SNN (T\textsubscript{8}) & MarkupLM & Triplet \\
    \bottomrule
  \end{tabular}
\end{table}

We evaluated eight configurations of \tool, combining four initial embedding models (Doc2Vec, BERT, ModernBERT, and MarkupLM) with two SNN network variants. All the configurations are listed in \autoref{table:eval-techniques} and are named SNN\textsubscript{1-8}. We trained an app-specific classifier for each of the nine web apps. In total, we trained and fine-tuned 72 SNN models for all nine applications. For each app, we used 80\% of the state pairs for training the classifier, 10\% for validation, and the remaining  10\% for testing.

\subsection{Procedure and Metrics}

\subsubsection{RQ\texorpdfstring{\textsubscript{1}}{1} (near-duplicate detection)}




We evaluated \tool's configurations towards maximizing near-duplicate detection by assessing them on the labeled $\mathcal{SS}$ dataset, and comparing them against baseline methods.
We used Precision, Recall, and the $F_1$-score, 
standard metrics for binary classification, where the positive class corresponds to \textit{near-duplicate} states.
Precision measures the proportion of correctly identified near-duplicate pairs among all predicted near-duplicates,
while Recall measures the proportion of actual near-duplicate pairs correctly identified. 
The $F_1$-score is their harmonic mean, providing a balanced view of both precision and recall.

\subsubsection{RQ\texorpdfstring{\textsubscript{2}}{2} (model \changed{quality})}

The crawl models contain redundant concrete states that Yandrapally et al.~\cite{2020-Yandrapally-ICSE} aggregated into the corresponding logical pages. 
Logical pages represent clusters of concrete pages that are semantically the same. To measure \tool's model \changed{quality} w.r.t. the ground truth, we computed the precision, recall, and $F_1$ scores, by simulating the crawling process using manual ground-truth classifications. In this context, Precision describes the ratio of unique states to the total number of states in the model, while Recall represents the ratio of covered functional bins to the total number of functional bins.



\subsubsection{RQ\texorpdfstring{\textsubscript{3}}{3} (code coverage)}

To assess the effectiveness of \tool when used for web testing, we crawled each web application in $\mathcal{SS}$ multiple times, each time varying the SAF. For all tools and apps, we set the same crawling time of 30 minutes. 
We used DANTE~\cite{biagiola2020dependency} to generate Selenium web test cases from the crawl sequences, execute the tests, and measured the web app code coverage. For JavaScript-based apps (Dimeshift, Pagekit, Phoenix, PetClinic), we measured \textit{client-side} code coverage using \texttt{cdp4j} (v. 3.0.8) library, i.e., the Java implementation of Chrome DevTools. For PHP-based apps (Claroline, Addressbook, PPMA, MRBS, MantisBT), we measured the \textit{server-side} code coverage using the \texttt{xdebug} (v. 2.2.4) PHP extension and \texttt{php-code-coverage} (v. 2.2.3).
During our evaluation, we first determined the maximum achievable code coverage through manual exploration. 

\subsubsection{RQ\texorpdfstring{\textsubscript{4}}{4} (time efficiency)}

Time efficiency is critical for any SAF integrated within a real‑time web‑crawling pipeline. Indeed, during crawling, thousands of state comparisons are performed, and even small delays can accumulate and significantly slow down exploration. To quantify this aspect, we measured the inference time of the investigated techniques, which directly impacts the efficiency and effectiveness of model inference and test generation.

We introduced two metrics related to time efficiency: model training time (RQ\textsubscript{4.1}) and inference time (RQ\textsubscript{4.2}). Specifically, model training time refers to the duration taken to train the SNN model. The inference time efficiency is evaluated using a subset of 1,000 randomly selected state pairs. Since FragGen, RTED, and PDiff methods are closely integrated within \crawljax, we extracted the distance calculation components from these methods and measured the execution time needed to generate the results.
\subsection{Results}

\subsubsection{RQ\texorpdfstring{\textsubscript{1}}{1} (near-duplicate detection)}

\autoref{table:rq1} presents the results for near-duplicate detection comparing the proposed technique against baseline methods. 
The configuration SNN (T\textsubscript{1}) achieves the best results, with the highest average F1-score of 0.96 for near-duplicate detection. When compared to the baselines, it provides a considerable improvement in near-duplicate detection, yielding approximately 23\%, 43\%, and 47\% increases in F$_1$-score compared to FragGen (0.76), RTED (0.67), and PDiff (0.65), respectively.

\begin{table}[t]
\centering
\caption{Near-Duplicate Detection (RQ\texorpdfstring{\textsubscript{1}})). Bold=best $F_1$.}
\setlength{\tabcolsep}{18.8pt}
\renewcommand{\arraystretch}{1.0}
\label{table:rq1}

\begin{tabular}{l c c c}
\toprule
\textbf{Technique} & \textbf{Precision} & \textbf{Recall} & $\mathbf{F}_1$ \\
\midrule
SNN (\(\text{T}_{1}\)) & \textbf{0.95} & \textbf{0.98} & \textbf{0.96} \\
SNN (\(\text{T}_{2}\)) & 0.68 & 0.97 & 0.77 \\
SNN (\(\text{T}_{3}\)) & 0.80 & 0.98 & 0.85 \\
SNN (\(\text{T}_{4}\)) & 0.58 & 0.93 & 0.70 \\
SNN (\(\text{T}_{5}\)) & 0.80 & 0.96 & 0.86 \\
SNN (\(\text{T}_{6}\)) & 0.60 & 0.90 & 0.70 \\
SNN (\(\text{T}_{7}\)) & 0.80 & 0.90 & 0.79 \\
SNN (\(\text{T}_{8}\)) & 0.58 & 0.93 & 0.69 \\
\midrule
FragGen                & 0.83 & 0.70 & 0.76 \\
RTED                   & 0.65 & 0.74 & 0.67 \\
PDiff                  & 0.72 & 0.62 & 0.65 \\
\bottomrule
\end{tabular}
\end{table}

\begin{figure}[t]
  \centering
  \includegraphics[width=.9\linewidth]{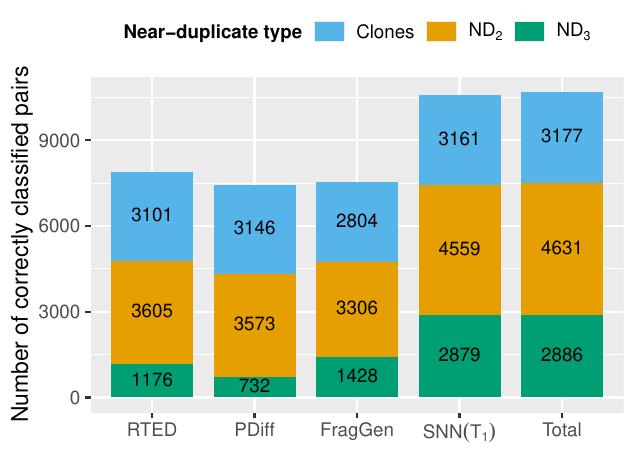}
  \caption[Correctly classified near duplicate state-pairs]{Correctly classified near-duplicate state-pairs.}
  \label{fig:state_pair_analysis}
\end{figure}

We further analyzed the number of correctly classified near-duplicate state pairs. To ensure a fair comparison with the baselines, we excluded App\textsubscript{4}, for which FragGen's results were not available.
Looking at \autoref{fig:state_pair_analysis}, SNN (T\textsubscript{1}) outperforms the baselines by a large margin on all categories: Clones, \( \text{ND}_{2} \) near-duplicates, and \( \text{ND}_{3} \) near-duplicates.
Overall, it correctly classifies $10{,}594$ pairs, outperforming FragGen ($7{,}538$), PDiff ($7{,}451$), and RTED ($7{,}882$), representing improvements of $41\%$, $42\%$, and $34\%$, respectively.

\begin{tcolorbox}[boxrule=0pt,frame hidden,sharp corners,enhanced,borderline north={1pt}{0pt}{black},borderline south={1pt}{0pt}{black},boxsep=2pt,left=2pt,right=2pt,top=2.5pt,bottom=2pt]
\textbf{RQ\textsubscript{1}}: \textit{
\tool achieves an overall improvement of 23\%---68\% of $F_1$ scores compared to all baseline methods. These enhancements correspond to an increased number of accurately classified near-duplicate state pairs.}
\end{tcolorbox}

\subsubsection{RQ\texorpdfstring{\textsubscript{2}}{2} (model quality)}

\autoref{table:rq2} shows the Precision (Pr), Recall (Re), and $F_1$ scores ($F_1$) of \tool and the baselines for all apps, along with the average per-app results. Although all \tool's configurations were evaluated, only the top-performing technique from RQ{\textsubscript{1}} (i.e., SNN($\text{T}_1$)) alongside the baseline is reported due to space limitations. 

Overall, \tool produces more accurate models (i.e., models more similar to the ground truth) than the competing techniques across all use cases. Specifically, the SNN($\text{T}_1$) configuration of \tool achieved the highest $F_1$-score of $0.81$, a substantial improvement of $62\%$ compared to FragGen, $35\%$ compared to RTED, and $59\%$ compared to PDiff. 

\begin{tcolorbox}[boxrule=0pt,frame hidden,sharp corners,enhanced,borderline north={1pt}{0pt}{black},borderline south={1pt}{0pt}{black},boxsep=2pt,left=2pt,right=2pt,top=2.5pt,bottom=2pt]
\textbf{RQ\textsubscript{2}}: \textit{
\tool shows an overall improvement of 35\% to 72\% over baseline methods and is better at approximating the ground-truth model than all existing techniques. 
}
\end{tcolorbox}

\subsubsection{RQ\texorpdfstring{\textsubscript{3}}{3} (code coverage)}

\begin{table}[t]
\centering
\caption{Model Quality (RQ\texorpdfstring{\textsubscript{2}})). Bold=best $F_1$.}
\setlength{\tabcolsep}{2.5pt}
\renewcommand{\arraystretch}{1.2}
\label{table:rq2}
    \begin{tabular}{
    l
    ccc
    ccc
    ccc
    ccc
    }
    \toprule
    & \multicolumn{3}{c}{\textbf{SNN (\(\text{T}_{1} \))}}
    & \multicolumn{3}{c}{\textbf{FragGen}} 
    & \multicolumn{3}{c}{\textbf{RTED}}
    & \multicolumn{3}{c}{\textbf{PDiff}} \\
    \cmidrule(lr){2-4} 
    \cmidrule(lr){5-7} 
    \cmidrule(lr){8-10}
    \cmidrule(lr){11-13} 
    \textbf{App}
    & \textbf{Pr} & \textbf{Re} & $\mathbf{F}_1$
    & \textbf{Pr} & \textbf{Re} & $\mathbf{F}_1$
    & \textbf{Pr} & \textbf{Re} & $\mathbf{F}_1$
    & \textbf{Pr} & \textbf{Re} & $\mathbf{F}_1$ \\
    \midrule
    \textbf{\(\text{App}_{1} \)} & 1.00 & 0.50 & 0.67 & 0.38 & 1.00 & 0.55 
    & 0.70 & 0.58 & 0.64 & 0.18 & 1.00 & 0.31 \\
    \textbf{\(\text{App}_{2} \)} & 0.62 & 1.00 & 0.77 & 0.92 & 0.92 & 0.92 
    & 1.00 & 0.83 & 0.91 & 0.94 & 0.81 & 0.86  \\
    \textbf{\(\text{App}_{3} \)} & 0.96 & 1.00 & 0.98 & 0.45 & 0.43 & 0.44 
    & 0.23 & 1.00 & 0.38 & 1.00 & 0.87 & 0.93 \\
    \textbf{\(\text{App}_{4} \)} & 0.65 & 0.93 & 0.76 & N/A & N/A & N/A
    & 0.53 & 0.64 & 0.58 & 0.20 & .64 & 0.31 \\
    \textbf{\(\text{App}_{5} \)} & 0.71 & 1.00 & 0.83 & 0.44 & 0.74 & 0.55 
    & 0.59 & 0.72 & 0.65 & 0.81 & 0.57 & 0.67 \\
    \textbf{\(\text{App}_{6} \)} & 1.00 & 0.67 & 0.80 & 0.10 & 0.67 & 0.18 
    & 0.80 & 0.76 & 0.78 & 0.35 & 0.81 & 0.49 \\
    \textbf{\(\text{App}_{7} \)} & 0.59 & 1.00 & 0.74 & 0.38 & 0.80 & 0.52 
    & 0.47 & 0.75 & 0.58 & 0.35 & 0.81 & 0.41 \\
    \textbf{\(\text{App}_{8} \)} & 1.00 & 0.70 & 0.82 & 0.30 & 0.80 & 0.43 
    & 0.55 & 0.60 & 0.57 & 0.09 & 0.90 & 0.17 \\
    \textbf{\(\text{App}_{9} \)} & 1.00 & 0.79 & 0.88 & 0.29 & 0.86 & 0.44 
    & 0.57 & 0.73 & 0.60 & 0.18 & 0.79 & 0.29 \\
    \midrule
    \textbf{Avg.} & \textbf{0.84} & \textbf{0.84} & \textbf{0.81} & 0.40 & 0.78 & 0.50 
    & 0.57 & 0.73 & 0.60 & 0.47 & 0.79 & 0.51 \\
    \bottomrule
    \end{tabular}
\end{table}

\autoref{table:rq3} presents the code coverage results for \tool as well as for the baseline methods.
Taking into account the average scores for all nine applications, \tool achieves the highest score compared to all baselines. Recall that the coverage results reported in \autoref{table:rq3} are normalized relative to the maximum achievable through manual exploration, which averaged at 38.4\% across all applications. 
\tool (SNN (T\textsubscript{1})) demonstrates the best code coverage of 71.99\% on average. This result indicates an improvement of approximately 6\%, 11\%, and 35\% over FragGen, RTED, and PDiff, respectively. 
A Wilcoxon signed-rank test and Vargha--Delaney effect size analysis were conducted over the nine applications. SNN (\(T_1\)) achieved the highest average coverage (71.99\%), outperforming FragGen (67.96\%), RTED (64.74\%), and PDiff (53.25\%). The improvements over RTED (\(p=0.019, A_{12}=0.83\)) and PDiff (\(p=0.019, A_{12}=0.89\)) are statistically significant with large effect sizes, while the difference over FragGen (\(p=0.078, A_{12}=0.72\)) shows a medium-to-large practical effect, though not significant at the 0.05 level. 

\begin{table}[t]
\centering
\caption{Code Coverage (RQ\texorpdfstring{\textsubscript{3}})). Bold=best average score.}
\label{table:rq3}
\setlength{\tabcolsep}{12.8pt}
\renewcommand{\arraystretch}{1.0}
    \begin{tabular}{
      l 
      c 
      c 
      c 
      c
    }
    \toprule
    \textbf{App}
      & \textbf{SNN (\(\text{T}_{1}\))}
      & \textbf{FragGen}
      & \textbf{RTED}
      & \textbf{PDiff} \\
    \midrule
    \textbf{\(\text{App}_{1}\)} & 87.89 & 91.48 
    & 90.11 & 44.98 \\
    \textbf{\(\text{App}_{2}\)} & 29.73 & 41.34 
    & 41.58 & 22.50 \\
    \textbf{\(\text{App}_{3}\)} & 85.10 & 67.17 
    & 47.87 & 48.75 \\
    \textbf{\(\text{App}_{4}\)} & 75.25 & 60.69 
    & 66.49 & 35.06 \\
    \textbf{\(\text{App}_{5}\)} & 35.76 & 38.18 
    & 32.06 & 23.94 \\
    \textbf{\(\text{App}_{6}\)} & 52.50 & 32.65 
    & 28.62 & 28.62 \\
    \textbf{\(\text{App}_{7}\)} & 95.18 & 96.46 
    & 92.36 & 92.31 \\
    \textbf{\(\text{App}_{8}\)} & 94.67 & 95.10 
    & 95.26 & 95.26 \\
    \textbf{\(\text{App}_{9}\)} & 87.81 & 88.60 
    & 88.28 & 87.81 \\
    \midrule
    \textbf{Avg.}              & \textbf{71.99} & 67.96 
    & 64.74 & 53.25 \\
    \bottomrule
    \end{tabular}
\end{table}

\begin{tcolorbox}[boxrule=0pt,frame hidden,sharp corners,enhanced,borderline north={1pt}{0pt}{black},borderline south={1pt}{0pt}{black},boxsep=2pt,left=2pt,right=2pt,top=2.5pt,bottom=2pt]
\textbf{RQ\textsubscript{3}}: \textit{
The tests generated from \tool crawl models achieve the highest code coverage scores, showcasing 6\%---21\% improvement thanks to the more accurate and complete web app models.
} 
\end{tcolorbox}

\subsubsection{RQ\texorpdfstring{\textsubscript{4}}{4} (time efficiency)}



\autoref{table:rq4_2} shows the average inference times for all techniques. 
\tool achieves the fastest inference time, which is an order of magnitude quicker than FragGen and PDiff, and significantly faster than RTED. This supports the applicability of \tool in real-world crawling scenarios, where thousands of state comparisons must be performed efficiently to keep exploration and test generation within practical time budgets.

\begin{table}[t]
\centering
\caption{Time Efficiency (RQ\texorpdfstring{\textsubscript{4}})) Inference time (Seconds). The best inference time is boldfaced.}
\setlength{\tabcolsep}{12pt}
\renewcommand{\arraystretch}{1.0}
\label{table:rq4_2}

\footnotesize{
    \begin{tabular}{lcc}
\toprule
\textbf{Technique}
 & \textbf{Avg. Inference Time} & \bf $\Delta$ vs. \tool \\
\midrule
\tool 
 & \bf 0.09  & -- \\
FragGen
 & 1.44  & $\times 16$\\
RTED
 & 0.68 & $\times 7.6$\\
PDiff
 & 2.03 & $\times 22.6$\\
\bottomrule
\end{tabular}
}
\end{table}

\begin{tcolorbox}[boxrule=0pt,frame hidden,sharp corners,enhanced,borderline north={1pt}{0pt}{black},borderline south={1pt}{0pt}{black},boxsep=2pt,left=2pt,right=2pt,top=2.5pt,bottom=2pt]
\textbf{RQ\textsubscript{4}}: \textit{\tool achieves the fastest inference time at $0.09$ seconds, delivering remarkable speedups of $16\times$ over FragGen, $7.6\times$ over RTED, and $22.6\times$ over PDiff. 
} 
\end{tcolorbox}



\subsection{Threats to Validity}\label{sec:ttv}

\subsubsection{Internal validity}
We compared all variants of \tool and baselines under identical experimental settings and on the same evaluation set (\autoref{sec:dataset}). \changed{In our experiments, a crawling time of 30 minutes allowed all crawls to explore all logical pages of the AUTs within the timeout. The test generation budget refers to the crawling time allowed for model inference, as the tests are extracted directly from the crawl sequences.}
The main threat to internal validity concerns our implementation of the testing scripts to evaluate the results, which we tested thoroughly.
The inference times reported in \autoref{table:rq4_2} compare the runtime of Java implementations of the baseline methods (FragGen, RTED, and PDiff within Crawljax) with Python implementations (\tool). Although all experiments were conducted on identical hardware, the startup overhead of the Java Virtual Machine and Python's  Global Interpreter Lock may introduce biases in the absolute timings.

\subsubsection{External validity}
The limited number of subjects in our evaluation poses a threat in terms of the generalizability of our results to other web apps. However, our evaluation considered nine representative web applications, that cover a range of architectures (single- and multi-page), GUI complexity, and dynamic behavior typical of modern SPAs. Results may however not transfer to applications whose technologies, languages, or interaction patterns were absent from our corpus.


\section{Related Work}\label{sec:related}

\subsection{End-to-End Web Test Automation}

Most E2E testing techniques rely on a web app model of the application under test, either manually generated by developers or automatically inferred, e.g., through crawling~\cite{mesbah:tse12,andrews2005testing,marchetto2008state,2015-Stocco-AST,2017-Stocco-SQJ,2016-Stocco-ICWE,2026-Karagoz-ICST}. 
Biagiola et al.~\cite{biagiola2017search, 2019-Biagiola-FSE-Diversity} use Page Objects to guide the generation of tests. Marchetto et al.~\cite{marchetto2008state} propose a combination of static and dynamic analysis to model the AUT into a finite state machine and generate tests based on multiple coverage criteria.
Mesbah et al.~\cite{mesbah:tse12} propose ATUSA, a tool that leverages the model of the AUT produced by \crawljax to automatically generate test cases to cover all the transitions of the model. 

These works do not address the redundancy in the web app model during crawling due to an ineffective SAF. These test generators can be used in conjunction with \tool to increase the accuracy of the inferred web app models.  

\subsection{Empirical Studies on Near-Duplicates}

Fetterly et al.~\cite{Fetterly} study the nature of near-duplicates during software evolution, reporting their low variability over time.
Yandrapally et al.~\cite{2020-Yandrapally-ICSE} compare different near-duplication detection algorithms as SAFs from the fields of computer vision and information retrieval. The paper reports on the challenge of finding an accurate SAF using DOM-based or visual techniques, which motivates the work in this paper to use advanced neural embeddings and classifiers from the deep learning domain. 

\subsection{Automated Near-Duplicate Detection}

Regarding detection of near-duplicates \textit{within} the same app, Crescenzi et al.~\cite{Crescenzi:2005:CWP:1086628.1086630} propose a structural abstraction for web pages and a clustering algorithm based on such abstraction. Di Lucca et al.~\cite{WESS01a,10.1109/CMPSAC.2002.1045051} evaluate the Levenshtein distance and the tag frequency for detecting near-duplicate web pages. Stocco et al.~\cite{2016-Stocco-ICWE} use clustering on structural features as a post-processing technique to discard near-duplicates in crawl models. 
Corazza et al.~\cite{corazza} propose the usage of tree kernels. 

Concerning detection of near-duplicates \textit{across} apps, researchers considered clustering techniques on raw structural features~\cite{Henzinger:2006:FNW:1148170.1148222,Manku:2007:DNW:1242572.1242592,Fetterly,Ramaswamy:2004:ADF:988672.988732,WESS01a,10.1109/CMPSAC.2002.1045051,Crescenzi:2005:CWP:1086628.1086630,2016-Stocco-ICWE}. 
Other works, such as the one by Henzinger~\cite{Henzinger:2006:FNW:1148170.1148222}, use shingles, i.e., $n$-grams composed of contiguous subsequences of tokens, to ascertain the similarity between web pages. 
Manku et al.~\cite{Manku:2007:DNW:1242572.1242592} use \texttt{simhash} to detect near-duplicates in the context of information retrieval, plagiarism, and spam detection. 
In this paper, we consider neural embeddings to train an SNN-based classifier for near-duplicate detection, and we illustrate that its usage for functional testing of web apps outperforms state-of-the-art techniques.

\subsection{Embeddings in Software Engineering}

Alon et al.~\cite{alon2019code2vec} present code2vec, a neural model for learning embeddings for source code, based on its representation as a set of paths in the abstract syntax tree. 
Hoang et al.~\cite{hoang2020cc2vec} propose CC2Vec, a neural network model that learns distributed representations of code changes. The model is applied for log message generation, bug fixing, patch identification, and just-in-time defect prediction.

Feng et al.~\cite{web2vec} use representation learning applied across web apps for phishing detection. 
Lugeon et al.~\cite{lugeon2022homepage2vec} propose Homepage2Vec, an embedding method for website classification. 
Namavar et al.~\cite{Namavar-EMSE} performed a large-scale experiment comparing different code representations to aid bug repair tasks. 
In this work, we use neural embeddings from the deep learning domain, a novel contribution in the context of automated crawling and web testing.

Ma et al.~\cite{GraphCode2Vec} propose GraphCode2Vec, a technique that joins code analysis and graph neural networks to learn lexically and program dependent features to support method name prediction. Dakhel et al.~\cite{dev2vec} propose dev2vec, an approach to embed developers' domain expertise within vectors for the automated assessment of developers' specialization. Jabbar et al.~\cite{test2vec} encode the test execution traces for test prioritization. 

Differently, we use several neural embeddings to train an SNN classifier that is used as SAF within a crawl-based test generator for functional testing.

\section{Conclusions and Future Work}\label{sec:conclusions}

In this paper, we aim to improve the crawlability of modern web applications by designing and evaluating \tool, a novel state abstraction function for web testing based on neural embeddings of web pages. 
Neural embeddings are used to train SNN-based classifiers for near-duplicate detection. We demonstrate their effectiveness in inferring accurate models for functional testing of web apps.
Our results show that crawl models produced with \tool have higher precision and recall than the ones produced with existing approaches. Moreover, these models allow test suites generated from them to achieve higher code coverage. 

Future work will explore richer embeddings to further enhance \tool's accuracy, including visual embeddings from web screenshots (e.g., via autoencoders~\cite{2020-Stocco-ICSE,2020-Stocco-GAUSS}), hybrid representations, and fine-tuned transformer models. In this study, we deliberately avoided fine-tuning to assess the intrinsic quality of web-aware embeddings and ensure fair, reproducible comparisons. We also plan to evaluate \tool's bug-finding capabilities, for instance, through mutation testing.
\ifCLASSOPTIONcaptionsoff
  \newpage
\fi



\balance
\bibliographystyle{ieeetr}
\bibliography{paper}
\end{document}